\documentclass[aps,prb,twocolumn,superscriptaddress]{revtex4-1}

\usepackage{amsmath,hyperref,color,amsthm,amssymb,bm}
\usepackage{graphicx}

\newcommand{\dd}[1]{\mathrm{d}#1\,}

\renewcommand{\vec}[1]{\bm{#1}}
\renewcommand{\Re}{\mathop{\mathrm{Re}}}

\DeclareMathOperator{\tr}{tr}
\DeclareMathOperator{\sech}{sech}

\DeclareMathOperator{\artanh}{artanh}

\newcommand{\sgn}{\mathop{\mathrm{sgn}}}

\begin{document}

\title{Quasiparticle entropy in superconductor/normal metal/superconductor
  proximity junctions in the diffusive limit}

\author{P. Virtanen}
\affiliation{NEST, Istituto Nanoscienze-CNR  and Scuola Normale Superiore, I-56127 Pisa, Italy}
\email{pauli.virtanen@nano.cnr.it}

\author{F. Vischi}
\affiliation{NEST, Istituto Nanoscienze-CNR  and Scuola Normale Superiore, I-56127 Pisa, Italy}
\affiliation{Dipartimento di Fisica, Universit\`{a} di Fisica, I-56127 Pisa, Italy}

\author{E. Strambini}
\affiliation{NEST, Istituto Nanoscienze-CNR  and Scuola Normale Superiore, I-56127 Pisa, Italy}

\author{M. Carrega}
\affiliation{NEST, Istituto Nanoscienze-CNR  and Scuola Normale Superiore, I-56127 Pisa, Italy}

\author{F. Giazotto}
\affiliation{NEST, Istituto Nanoscienze-CNR  and Scuola Normale Superiore, I-56127 Pisa, Italy}

\begin{abstract}
  We discuss the quasiparticle entropy and heat capacity of a dirty
  superconductor-normal metal-superconductor junction. In the case of
  short junctions, the inverse proximity effect extending in the
  superconducting banks plays a crucial role in determining the
  thermodynamic quantities. In this case, commonly used approximations
  can violate thermodynamic relations between supercurrent and
  quasiparticle entropy. We provide analytical and numerical results
  as a function of different geometrical parameters. Quantitative
  estimates for the heat capacity can be
  relevant for the design of caloritronic devices or radiation sensor
  applications.
\end{abstract}

\maketitle

\section{Introduction}
\label{sec:intro}

Recently a growing interest has been put on the investigation of
thermodynamic properties of nanosystems, where coherent effects can be
both of fundamental interest and useful for applications.
\cite{esposito_2009,campisi_2011,giazotto_2006,carrega_2016,fornieri_2016b}
In particular, superconductor junction systems have attracted
interest, as they exhibit phase-dependent thermal transport enabling
coherent caloritronic devices,
\cite{giazotto_2006,giazotto2012-jhi,strambini_2014,giazotto2012-pcs,martinez-perez2013-ept,fornieri2016-npe,paolucci_2016}
and have properties useful for cooling systems in solid-state devices
\cite{muhonen_2012,solinas_2016,nguyen_2016,courtois_2016}. Conversely,
they enable conversion between thermal currents and electric signals,
leading to applications in electronic thermometry
\cite{saira_2016,giazotto_2006,feshchenko_2015} and bolometric sensors
and single-photon detectors
\cite{wei_2008,govenius_2016,semenov_2002,engel_2015,voutilainen_2010,giazotto2008-upj,govenius2014-mnb,govenius2016-dzm,karasik_2012}.
In such applications, detailed understanding of the thermodynamic
aspects of hybrid superconducting--normal metal structures is crucial,
in particular, the interplay between the energy and entropy related to
quasiparticles and supercurrents.

The entropy $S$ of noninteracting quasiparticles at equilibrium is
generally determined by their density of states (DOS). In the
superconducting state, it is modified by the appearance of an energy
gap in the spectrum.  In extended Josephson junctions such as
superconductor--normal metal--superconductor (SNS) structures, the
modification of the DOS depends both on the formation of Andreev bound
states inside the junction and the inverse proximity effect in the
superconducting banks, both being modulated by the phase difference
$\varphi$ between the superconducting order parameters.
\cite{likharev79,pannetier_2000} Reflecting the fact that the Andreev
bound states carry the supercurrent $I$ across the junction, a
thermodynamic Maxwell relation
\begin{equation}
  \frac{\dd{{S}}}{\dd{\varphi}}=-\frac{\hbar}{2e}\frac{\dd{I}}{\dd{T}}
  =
  -\frac{\dd{^2 F}}{\dd{T}\dd{\varphi}}
  \label{eq:TD_relation}
\end{equation}
connects the entropy and the supercurrent to the temperature $T$ and
phase derivative of the free energy $F$.  The entropy in
superconductors can be expressed in terms of the DOS
\cite{bardeen1957-tos} or in terms of Green functions
\cite{gorkov59b,eilenberger1968-tog}. Moreover, the phase-dependent
part of ${S}$ can be obtained from the current-phase relation
$I(T,\varphi)$, \cite{likharev79,golubov_2004}, by applying
Eq.~\eqref{eq:TD_relation}, a contribution important in short
junctions
\cite{beenakker1991-spc,beenakker1991-jct,beenakker1991-spc}.  The
different expressions are mathematically equivalent (see e.g.
Refs.~\onlinecite{kosztin1998-feo,kos1999-gef}). Such equivalences
however can be broken by approximations: in particular, the ``rigid
boundary condition'' approximation \cite{likharev79,golubov_2004}, in
which the inverse proximity effect in the superconductors is
neglected, invalidates DOS-based expressions for entropy. Although
such approximations are appropriate for many purposes, they can give
wrong results for thermodynamic quantities when boundary effects matter.

Heat capacity \cite{fulde1967,zaitlin1982-hcd,kobes1988-fec} and free
energy boundary contributions
\cite{hu1972-tse,eilenberger1975-beb,blackburn1975,kos1999-gef} in
NS systems were considered in several previous works; also
experimentally, \cite{lechevet1972-tpe,manuel1976-shj} close to the
critical temperature $T_c$. The inverse proximity effect in the
superconducting banks of diffusive NS structures is also well studied.
\cite{kupriyanov1982-tpe,likharev79,blackburn1975,golubov_2004}
The entropy and heat capacity in diffusive SNS junctions
were discussed in Refs.~\onlinecite{rabani2008-pde,rabani_2009}, but
neglecting the inverse proximity effect, which limits the validity of
the results to long junctions only.

In this work, we discuss the proximity effect contributions to the
entropy and heat capacity in SNS structures of varying size. We also
point out reasons for the discrepancies that appear with the rigid
boundary condition approximation in the quasiclassical formalism. We
provide analytical results for limiting cases, and discuss the
cross-over regions numerically.

The paper is organized as follows. In Sec.~\ref{sec:theory} we
introduce the theoretical formalism, based on the Usadel equations,
and all basic definitions.  In Sec.~\ref{sec:Maxwell} we discuss the
origin of inconsistencies in the rigid boundary condition
approximation. In Sec. \ref{sec:inverse_entropy} we present
quantitative results for the entropy inside the inverse proximity
region and the total entropy. We also show results for the heat
capacity in Sec.~\ref{sec:heatcapacity} and the effect of inverse proximity contributions on this
quantity. Sec.\ref{sec:concl} concludes with discussion.

\section{Model and basic definitions}
\label{sec:theory}

\begin{figure}
  \includegraphics{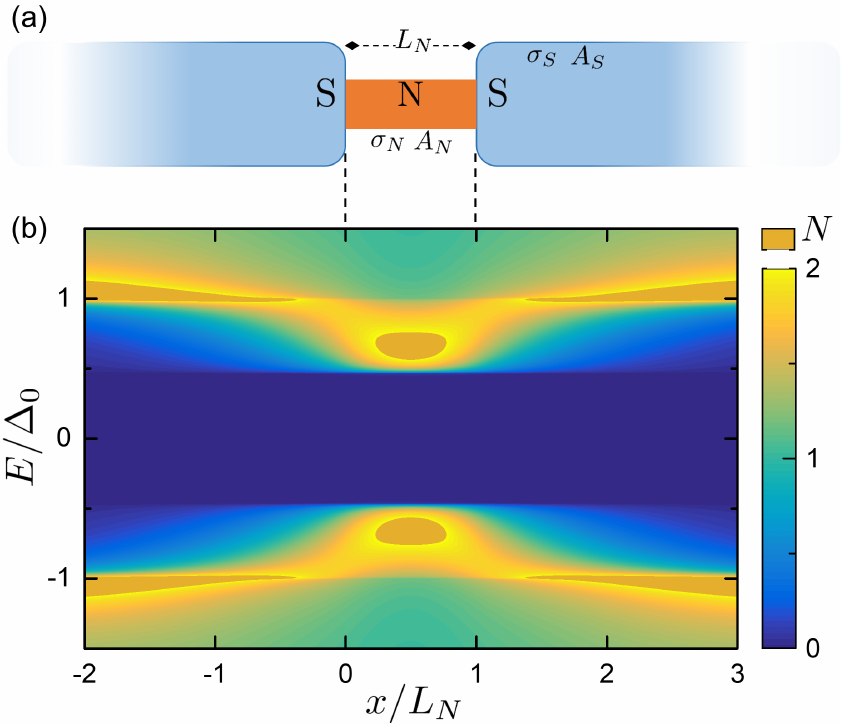}
  \caption{(a)
    Schematic of a SNS junction consisting of two superconducting (S) leads
    in clean electric contact with a normal (N) diffusive nanowire of length
    $L_N$. The S and N parts have cross sections $A_{S,N}$ and conductivity
    $\sigma_{S,N}$. (b) Normalized density of states (DOS) $N(E,x)$
    for $\sigma _S A_S/\sigma_N A_N=1$, $L_N/\xi_{N}=1$ and phase difference
    $\varphi=0$.
  }
  \label{fig:schematics}
\end{figure}

Here we consider a Josephson junction as schematically depicted in
Fig.\ref{fig:schematics}(a), where two superconducting banks (S) are
in clean electric contact with a normal (N) diffusive wire of length
$L_N$. The S and N parts are characterized by cross sections $A_{S,N}$
and electrical conductivities $\sigma _{S,N}$,
respectively. Microscopically, the two diffusive regions are
characterized by the diffusion coefficients $D_{S,N}$ and density of
states (DOS) per spin $\mathcal{N}_{0,S}$ and $\mathcal{N}_{0,N}$ at
Fermi level. These quantities are related to conductivities via
$\sigma _j = 2 e^2 D_j \mathcal{N}_{0,j}$, where, the factor 2 takes
into account spin degeneracy.

The presence of superconducting leads induces superconducting
correlations in the electrons in the normal metal. The correlations at
energy $E$ are associated with a characteristic coherence length
$\xi_E$, which in general may differ from the superconducting
coherence length $\xi_{N/S}\equiv\sqrt{\hbar{}D_{N/S}/|\Delta|}$. The
superconductors have order parameter $\Delta$, with phase difference
$\varphi$ across the junction. We also assume that the superconductor
material has critical temperature $T_c$ in bulk.

The entropy density ${\cal S}$, and thus the total entropy
$S(T, \varphi) =\int{}dx{\cal S}(x,T,\varphi)$, can be written in
terms of the quasiparticle spectrum:
\begin{multline}
  \label{eq:dosentropy}
  {\cal S}(x,T,\varphi)
  = \\=
  -
  4 {\cal N}_0
  \int_{-\infty}^\infty\dd{E}
  N(E,x,\varphi)f(E, T)\ln{f(E,T)}
  \,,
\end{multline}
where $N(E,x,\varphi)$ is the (reduced) local density of states and
$f(E, T)=1/(e^{E/T} + 1)$ the Fermi distribution function. The
normal-state result without proximity effect is found by setting
$N(E,x,\varphi)=1$ in the above expression, giving
$\mathcal{S}_n(T)= 2\pi^3{\cal N}_0T/3$.  The entropy density
${\cal S}(x,T,\varphi)$ can also be written as:
\begin{equation}
  {\cal S}(x,T, \varphi) = {\cal S}_n(x,T) -
  \frac{\dd{\mathcal{F}_S(x,T,\varphi)}}{\dd{T}}
  \,,
\end{equation}
where ${\cal F}_S=\mathcal{F}-\mathcal{F}_n$ is the difference in the free energy density
between superconducting and normal states.

A functional for the free energy density difference can be expressed
in terms of isotropic quasiclassical Green functions $\hat{g}$ in the
dirty limit:
\cite{eilenberger1968-tog,usadel1970-gde,altland1998-ftm,taras-semchuk2001-qif}
\begin{align}
  \label{eq:eilenberger}
  {\cal F}_S
  &=
  {\cal N}_0
  |\Delta|^2
  \ln\frac{T}{T_c}
  + 
  \pi{}T
  {\cal N}_0\sum_{\omega_n}[\frac{|\Delta|^2}{\omega_n} + \mathcal{L}(i\omega_n)]
  \,,
  \\
  \mathcal{L} &=
  \tr\{
  \omega_n[\sgn(\omega_n) - \tau_3\hat{g}]
  -
  (\Delta\tau_++\Delta^*\tau_-)\hat{g}
  +
  \frac{D}{4}(\hat{\nabla}\hat{g})^2
  \}
  \,,
\end{align}
where $\tau_j$ indicate Pauli matrices in the Nambu space. The above
expression assumes the quasiclassical constraint $\hat{g}^2=1$.  The
long gradient $\hat{\nabla}X=\nabla{}X-i[\vec{A}\tau_3,X]$ contains
the vector potential.  The superconducting order parameter is
$\Delta=|\Delta|e^{i\phi}$ and $\omega_n=2\pi{}T(n+\frac{1}{2})$ are
Matsubara frequencies. The reduced density of states reads
$N(E,x,\varphi)=\frac{1}{2}\Re\tr\tau_3\hat{g}(E+i0^+,x,\varphi)$.
Here and below, $e=\hbar=k_B=1$, unless otherwise specified.

The quasiclassical functions can be determined by the Usadel equation,
\cite{usadel1970-gde} which is an Euler-Lagrange equation
$\frac{\delta{}F}{\delta{\hat{g}}}=0$ for free energy
$F=\int\dd{x}\mathcal{F}$, under the constraint $\hat{g}^2=1$.
Explicitly we have
\begin{equation}
\label{eq:usadel}
D\hat\nabla\cdot(\hat{g}\hat{\nabla}\hat{g})-[\omega_n\tau_3+\Delta\tau_++\Delta^*\tau_-,\hat{g}]=0\,,
\end{equation}
The supercurrent $I$ along the $x$-axis, at a given position $x_0$,
can be expressed in terms of the above functional as
\begin{align}
  \label{eq:supercurrent}
  I(x_0)
  =
  \frac{\delta{}F}{\delta A_x(x_0)}
  =
  \frac{2e}{\hbar}\frac{\dd{F}}{\dd{\varphi}}
  \,.
\end{align}
Note that this quantity is generally conserved only if the order
parameter $\Delta$ is self-consistent,
$\delta{F}/\delta\Delta=\delta{F}/\delta\Delta^*=0$.
\footnote{
  As usual, the complex conjugate is formally a separate variable
  in the derivative.
}
If this is not
the case, the equalities in Eq.~\eqref{eq:supercurrent} remain valid
if the derivative vs. $\varphi$ is understood to be taken with
respect to the order parameter phases as
$\phi(x,\varphi)=\phi_0(x)+\theta(x-x_0)\varphi/2-\theta(x_0-x)\varphi/2$.

From Eq.~\eqref{eq:TD_relation} and known current-phase relations
\cite{golubov_2004}, the entropy associated to Andreev bound states
can also be obtained up to a $\varphi$-independent term. From the result
relevant for short junctions in the diffusive limit, \cite{kulik1975-cmt}
\begin{gather}
  \label{eq:KOI}
  I(T,\varphi)
  = \frac{4\pi T}{e R_N}
  \sum_{\omega_n}
  \frac{\Delta \cos(\varphi/2)}{\Omega_n}
  \mathrm{arctan}\frac{\Delta \sin(\varphi/2)}{\Omega_n}
  \\
  \label{eq:KOentropy}
  \begin{split}
  S(T,\varphi)-S(T,\varphi=0)
  =
  -\frac{\hbar}{2e}
  \int_0^{\varphi}\dd{\varphi'}
  \frac{\dd{I}}{\dd{T}}
  =
  \\
  =
  \frac{\pi\hbar}{2e^2R_NT^2}
  \int_{|\Delta||\cos\frac{\varphi}{2}|}^{|\Delta|}
  \dd{E}E
  \sech^2\Bigl(\frac{E}{2T}\Bigr)
  \\
  \times
  \ln
  \frac{
    |\Delta||\sin\frac{\varphi}{2}| + \sqrt{E^2 - |\Delta|^2\cos^2\frac{\varphi}{2}}
  }{
    \sqrt{|\Delta|^2 - E^2}
  }
  \,,
  \end{split}
\end{gather}
where $R_N=L_N/(\sigma_N A_N)$ is the resistance of the normal region
and $\Omega_n^2=\omega_n^2 + |\Delta|^2 \cos^2(\varphi/2)$. The
temperature dependence of $\Delta(T)$ is ignored, which is valid at
low temperatures.  The $\varphi=0$ term can be determined to be
$S(\varphi=0)=0$ (see below). This result ignores the inverse
proximity effect --- qualitatively, including it would result to an
increase of $L_N$ by a multiple of the coherence
length. \cite{likharev79}

For simplicity, in the following we assume transparent SN
interfaces, described by the quasi 1D boundary conditions (e.g. at the
left SN contact $x=0$), \cite{kupriyanov1988-iob,nazarov_1994}
\begin{equation}
  \hat{g}\rvert_{x\rightarrow 0^-}
  =
  \hat{g}\rvert_{x\rightarrow 0^+}
  \,,
  \;
  \sigma_S A_S \hat{g} \partial _x \hat{g} \rvert_{x\rightarrow 0^-} = \sigma_N A_N \hat{g} \partial _x \hat{g} \rvert_{x\rightarrow 0^+}
  \,.
  \label{eq:boundary1}
\end{equation}
and similarly on the right SN interface at $x=L_N$. The
cross-sectional areas appear in the above equations from conservation
of the matrix current $\hat{g}\nabla\hat{g}$; \cite{nazarov_1994} for
$A_S\ne{}A_N$ such quasi-1D approximation ignores details of the
current distribution at the contact, which requires that the
cross-sectional size is small compared to superconducting coherence
length $\xi_{S/N}$.

The rigid NS boundary condition approximation is formally given by the
limit $A_S\sigma_S\to\infty$, where there is no inverse proximity
effect. In this case, the Green function inside S approaches its bulk
value, and the boundary conditions are replaced by
$\hat{g}\rvert_{x=0,L_N}=\hat{g}\rvert_{S,\mathrm{BCS}}$.

For reference, we show in Fig. \ref{fig:schematics}(b) the behavior of
the density of states $N(E, x,\varphi)$ at $\varphi=0$, computed
numerically from $\hat{g}$ using the above approach.  The result assumes a
non-self-consistent $\Delta(x)=|\Delta|$ in the S regions.  Far from
the N region ($x\to\pm\infty$), the DOS approaches the BCS form
with energy gap $|\Delta|$, and towards the N region a minigap $E_g$
\cite{zhou_1998} becomes clearly visible.

\section{Rigid boundary conditions}
\label{sec:Maxwell}

Supercurrent and entropy are connected by an exact
Maxwell relation:
\begin{equation}
  \frac{\partial I}{\partial T} = -\frac{2 e}{\hbar} \frac{\partial S}{\partial \varphi}
  \label{eq:maxwell2}
\end{equation}
This relation does not hold between Eqs.~\eqref{eq:dosentropy}
and~\eqref{eq:supercurrent} within the rigid boundary condition
approximation, as one can argue directly as follows. Within the
approximation, the phase dependent part of the entropy $S$ is
localized in the N region; hence, the volume integral of
Eq.~\eqref{eq:dosentropy} scales as $\partial_\varphi S \propto
L_N$. On the other hand, the supercurrent~\eqref{eq:KOI} obtained
under the same approximation scales as
$\frac{\dd{I}}{\dd{T}}\propto{}L_N^{-1}$. Therefore one immediately
recognizes that the left and right-hand sides of
Eq.~\eqref{eq:maxwell2} have different dependence on $L_N$,
demonstrating the inconsistency between
supercurrent and entropy within the approximation.

\begin{figure}
  \includegraphics{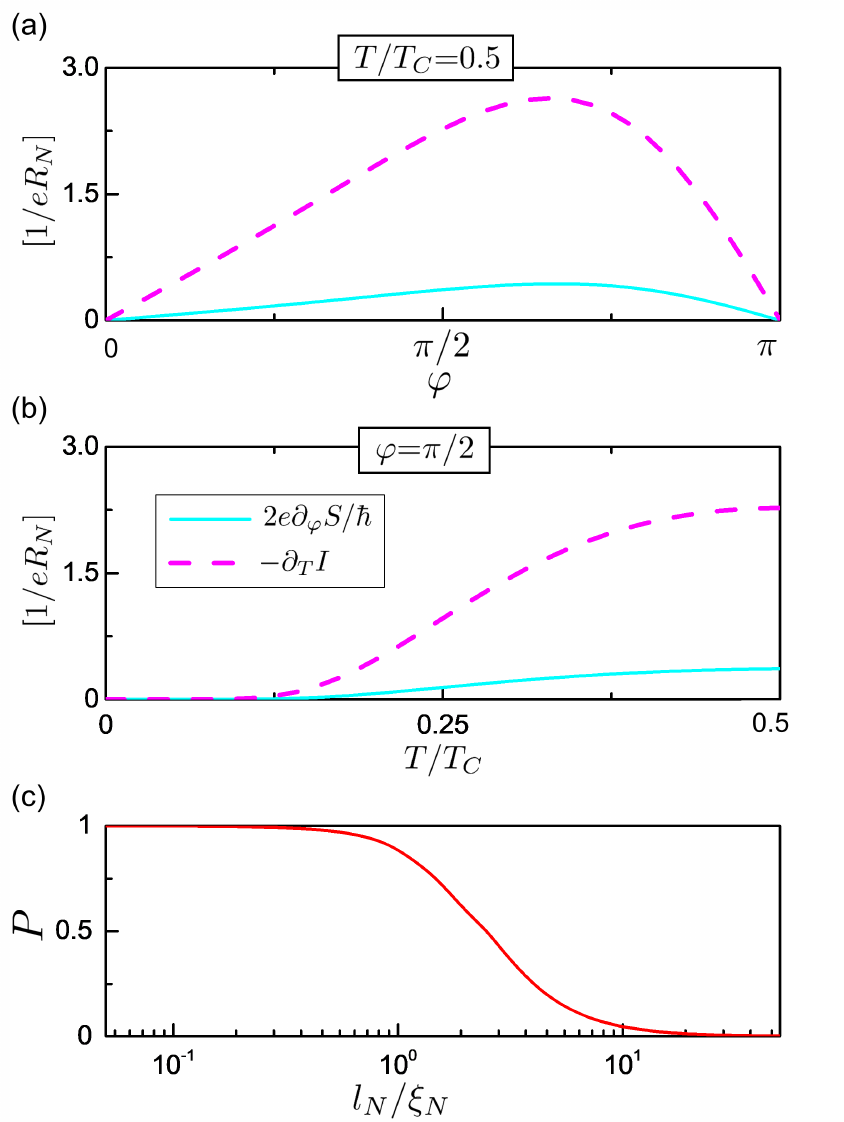}
  \caption{Inconsistency of the Maxwell relation with
    the rigid boundary condition approximation, using
    Eqs.~\eqref{eq:dosentropy},\eqref{eq:supercurrent}. (a)
    Left and right-hand sides of Eq.~\eqref{eq:maxwell2} vs.
    $\varphi$,
    at fixed temperature $T=0.5T_c$, for $\sigma_SA_S=\sigma_NA_N$
    and $L_N=\xi_N$.
    (b) Same vs. temperature at $\varphi=\pi/2$.
    (c) The relative discrepancy $P$ (see text)
    as a function of the normal region size $L_N/\xi_N$.
  }
\label{fig:discrepancy}
\end{figure}

The magnitude of the discrepancy in the rigid boundary condition
approximation is shown in Fig.~\ref{fig:discrepancy}(a,b), showing the
that the left and right-hand sides of Eq.~\eqref{eq:maxwell2} do not
match, as functions of phase difference $\varphi$ and temperature
$T$. Fig.~\ref{fig:discrepancy}(c) shows the dependence on $L_N/\xi_N$
of the relative discrepancy
\begin{equation}
  P=\max _{(\varphi,T)} \left|\frac{\partial _T I +
    \frac{2 e}{\hbar} \partial_\varphi S}{\partial _T I} \right|
  \,.
\end{equation}
It decreases with increasing junction length $L_N$, and remains
significant up to $L_N$ several times the coherence length $\xi_N$.
As one would expect, the discrepancy becomes negligible for long
junctions ($L_N\gg \xi_N$).

Let us now point out a mathematical relation between
Eqs.~\eqref{eq:dosentropy} and~\eqref{eq:eilenberger} related to the
discrepancy. Consider a modified Eilenberger functional,
$\mathcal{L}_{\zeta}=\mathcal{L}\rvert_{
  \omega_n\mapsto\omega_n+i\zeta}$, where the $\omega_n$ appearing
explicitly in $\mathcal{L}$ are replaced by $\omega_n+i\zeta$,
(cf. Ref.~\onlinecite{burkhardt1994-ffl}) and define the corresponding
Green functions $\hat{g}_{\zeta}$ satisfying
$\delta{}{\cal F}/\delta\hat{g}\rvert_{\hat{g}_\zeta}=0$
and keep $\Delta$ fixed. Recall that
the analytic continuation of the sign function is given by
$\sgn{}z=z/\sqrt{z^2}=\sgn\Re{}z$.  The stationary value of the
functional then satisfies for real $\zeta$
\begin{multline}
  \frac{\dd{}}{\dd{\zeta}}\mathcal{F}_{S,\zeta}\rvert_{\hat{g}_\zeta,\Delta}
  =
  \pi{}T{\cal N}_0\sum_{\omega_n}\tr[\sgn(\omega_n)-\tau_3\hat{g}_\zeta(i\omega_n)]
  \\
  =
  -{\cal N}_0
  \int_{-\infty}^\infty\dd{E}[N_\zeta(E) - 1]
  \tanh\frac{E}{2T}
  \,.
\end{multline}
The second line follows by standard analytic continuation, where
$N_\zeta(E)=\frac{1}{4}\tr\tau_3[g_\zeta(E+i0^+)-g_\zeta(E-i0^+)]$.
Suppose now that the boundary conditions are \emph{energy-indepenent},
i.e., invariant under transformation
$\omega_n\mapsto{}\omega_n+i\zeta$ of explicit frequency arguments: in
this case $\hat{g}_\zeta(i\omega_n)=\hat{g}(i\omega_n-\zeta)$ and
$N_\zeta(E)=N(E-\zeta)$ coincide with the energy-shifted Green
function and the corresponding DOS.  It is worth to notice that
$\mathcal{F}_{S,\zeta}\to\text{const}(T)$ for $\zeta\to\infty$ while
$\hat{g}(i\omega_n-\zeta)\to\tau_3\sgn(\omega_n)$. Moreover,
recalling the relation
\begin{multline}
\int_{-\infty}^0\dd{\zeta}\frac{\dd}{\dd{T}}\tanh\frac{E+\zeta}{2T}\\
=-2[f(E,T)\ln{}f(E,T)
+ (1-f(E,T))\ln(1-f(E,T))]~,
\end{multline}
it follows that $\partial_T\mathcal{F}_{S}=-\mathcal{S}_S$.  Finally,
setting $\Delta$ to its self-consistent value (which is a saddle point
of $\mathcal{F}_{S}$), we find Eqs.~\eqref{eq:dosentropy}
and~\eqref{eq:eilenberger} are equivalent, under the assumption that
the boundary conditions do not depend on energy.

The boundary value $\hat{g}_{S,\mathrm{BCS}}(i\omega_n)$ however is
strongly energy dependent, which breaks the above argument and causes
the discrepancy between Eqs.~\eqref{eq:dosentropy}
and~\eqref{eq:eilenberger},\eqref{eq:supercurrent}.  It is interesting
to note that a similar issue does not occur in an NSN structure under
an analogous approximation (also inspected numerically; not shown),
because in that case the value $\hat{g}_{N}=\tau_3\sgn\omega_n$
imposed in the boundary condition is invariant under
$\omega_n\mapsto{}\omega_n+i\zeta$.  This happens also for insulating
interfaces ($\hat{n}\cdot\hat{\nabla}\hat{g}=0$) or for periodic
boundary conditions, which are functionals of $\hat{g}$ with no
explicit dependence on $\omega_n$.

\begin{figure}{}
  \includegraphics{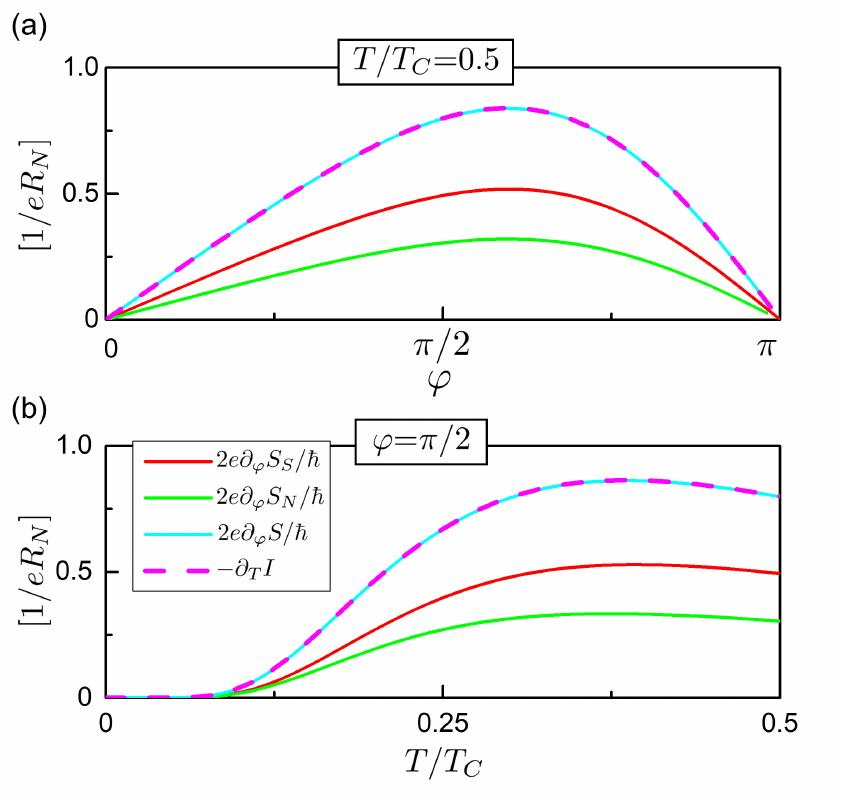}
  \caption{
    Maxwell relation including the inverse proximity effect.
    (a)
    Left and right-hand sides of Eq.~\eqref{eq:maxwell2} vs.
    $\varphi$, for $T=0.5T_c$, $\sigma_SA_S=\sigma_NA_N$, $L_N=\xi_N$.
    The entropy contributions from the N and S regions,
    $S=S_S+S_N$, are also shown separately.
    (b)
    Same vs. temperature at $\varphi=\pi/2$.
  }
  \label{fig:agreement}
\end{figure}

The apparent thermodynamic discrepancy can be eliminated by properly
taking into account the inverse proximity effect. For example,
replacing the rigid superconducting terminals by S wires of length
$L_S$. Below, we adopt a S'SNSS' geometry, with the boundary
conditions
$\hat{g}\rvert_{x=-L_S}=\hat{g}\rvert_{x=L_N+L_S}
=\hat{g}\rvert_{\mathrm{BCS}}$.  The effect of the boundary values is
rapidly suppressed and vanishes in the limit $L_S\to\infty$.

We show results for such SS'NS'S structure in
Fig.~\ref{fig:agreement}. In them, the
Maxwell relation~\eqref{eq:maxwell2} applies for any $L_N$. For
simplicity, this calculation does not use a self-consistent $\Delta$,
so that the phase derivative is to be understood as explained below
Eq.~\eqref{eq:supercurrent}.  Note that the entropy contribution from
the superconductor regions dominates for the parameters chosen.

\section{Inverse proximity effect}
\label{sec:inverse_entropy}

Let us consider the inverse proximity effect in more detail.  We
define the entropy difference $\delta S_S$ due to the inverse
proximity effect in the superconducting region as:
\begin{multline}
  \delta S_S = S_S - S_{{\rm BCS}}=
  \\
  =
  -4\int^\infty_{-\infty}\dd{E} \int_S\dd{x} {\cal N}_{0,S} \delta N (E, x,\varphi) f(E,T) \ln f(E,T)
  \,,
\end{multline}
where ${S}_{{\rm BCS}}$ is the entropy of a bulk BCS superconductor
and $\delta N (E, x,\varphi) = N(E,x,\varphi)-N_{{\rm BCS}}(E)$ is the
difference of the local density of states from the BCS expression.
Moreover, we define dimensionless parameters
\begin{equation}
  a
  =
  \sigma_SA_S/(\sigma_NA_N)
  \,,
  \qquad
  \ell
  =
  L_N/\xi_N
\end{equation}
for the discussion below.

Analytical solutions can be obtained in the limiting cases of
short junction $\ell\ll1$ at phase differences $\varphi=0$ and
$\varphi=\pi$. A solution to the Usadel equation in a semi-infinite
superconducting wire with uniform $\Delta=\pm|\Delta|$ is given by
\begin{equation}
\hat{g}=\tau_3\cosh\theta + i\tau_2\sinh\theta
\end{equation}
where (cf. Ref.~\onlinecite{zaikin81})
\begin{align}
  \theta(x) = \theta_S - 4\artanh\Bigl(e^{-\sqrt{2}(x-L_N)/\xi_E}\tanh\frac{\theta_S-\theta(L_N)}{4}\Bigr)
  \,,
\end{align}
and $\xi_E=(1 - E^2/|\Delta|^2)^{-1/4}\xi_N$ and
$\theta_S=\artanh\frac{|\Delta|}{E+i0^+}$.
The spatially integrated change
in the superconductor DOS can be evaluated based on this solution:
\begin{multline}
  \int_S\dd{x}\delta N(x,E) =\\
  = 
  \sqrt{2}\Re\Bigl[
  \xi_E
  \cosh\theta_S\bigl(\cosh\frac{\theta_S-\theta(L_N)}{2}-1\bigr)+
  \notag \qquad
  \\-
  \xi_E
  \sinh\theta_S\sinh\frac{\theta_S-\theta(L_N)}{2}
  \Bigl]
\end{multline}
For $L_N\ll\xi_E$, the Usadel equation in the N region can be
approximated as $\partial_x^2\theta(x)=0$.  Matching to the boundary
condition
$\sigma_NA_N\partial_x\theta_{N}=\sigma_SA_S\partial_x\theta_{S}$ at
the two NS interfaces results to
\begin{align}
  \theta(L_N)
  =
  \begin{cases}
    \theta_S & \text{for $\varphi=0$,}
    \\
    \sqrt{2}\frac{\xi_E}{\xi_N} a \ell \sinh\frac{\theta_S-\theta(L_N)}{2}
    &
    \text{for $\varphi=\pi$}
    \,,
  \end{cases}
  \label{eq:phase_discontinuity}
\end{align}
from which $\theta(L_N)$ can be solved. For the entropy at
$\varphi=0$, this gives a trivial solution $\delta S_S = 0$.  On the
other hand, at $\varphi=\pi$, we have for temperatures $T\ll|\Delta|$,
\begin{align}
  \delta{} S_S(\varphi=\pi)
  \simeq
  \frac{4\pi^2}{3} T
  {\cal N}_{0,S} A_S \xi_N
  \times
  \begin{cases}
    1 \,, & \text{for $a\ell\ll1$,}
    \\
    \frac{\pi}{2a\ell} \,, & \text{for $a\ell\gg1$.}
  \end{cases}
  \label{eq:limits}
\end{align}
The full temperature dependence for $\ell\to0$ reads
\begin{multline}
  \label{eq:ST-pi-analytic}
  \delta{} S_S(\varphi=\pi)
  =
  -\frac{16}{\sqrt{2}}
  {\cal N}_{0,S} A_S
  \int_{-\infty}^\infty\dd{E}
  f(E,T)\ln f(E,T)
  \\\quad\times
  \Re[(\cosh\frac{\theta_S}{2} - \cosh\theta_S)\xi_E]
  \,.
\end{multline}
For cross-over regions, the boundary condition matching would need to
be solved numerically.

\begin{figure}
  \includegraphics{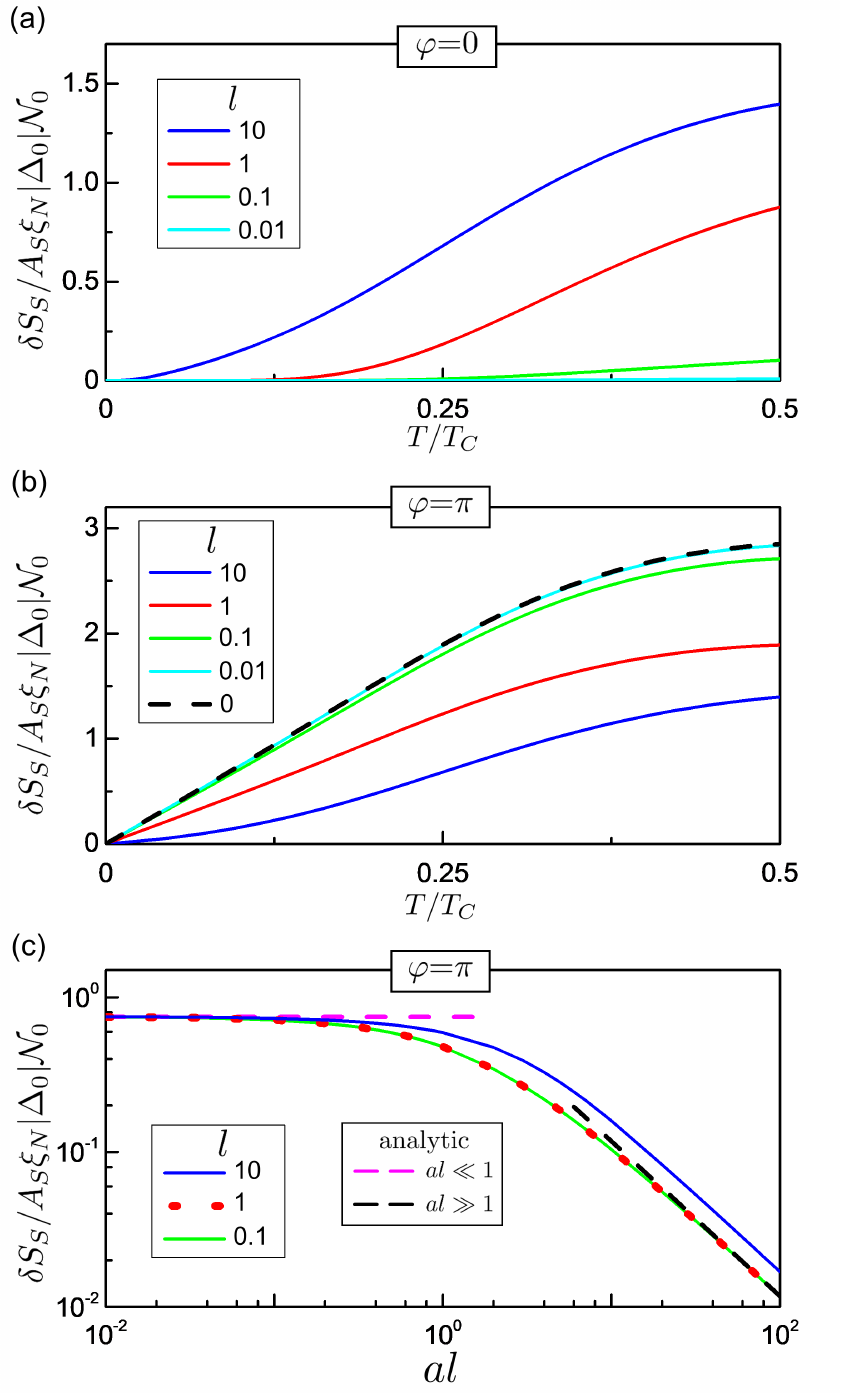}
  \caption{
    Behavior of the entropy variation $\delta S_S$ of the superconducting leads.
    (a)
    Temperature dependence at $\varphi=0$ for $a=1$ and different $\ell$.
    (b)
    Same at $\varphi=\pi$. Result~\eqref{eq:ST-pi-analytic} for $\ell=0$
    is also shown (dashed).
    (c)
    Dependence of $\delta  S_S(\varphi=\pi)$ on $\ell$ and $a$, at $T/T_C=0.1$.
    Limiting behavior from Eq.~\eqref{eq:limits} is indicated (dashed).
  }
  \label{fig:entropy_limits}
\end{figure}

The behavior in the rigid boundary condition limit (i.e. $a\to\infty$)
can be understood based on the above result.  For the entropy, the
short-junction rigid-boundary limit $a\to\infty$, $\ell\to0$ is not
unique, but results depend on the product $a\ell$.  Generally, the
entropy is proportional to $\hbar/(e^2R_{\rm tot})$, where
$R_{\rm tot}$ is the resistance of $\xi_S$-length superconductor
segment in series with the normal wire, as can be expected a priori
\cite{likharev79}.

Figures~\ref{fig:entropy_limits}(a,b) show the geometry dependence of
the proximity effect contribution $\delta{S}_S$ to the entropy, for
$\varphi=0$ and $\varphi=\pi$. Generally, $\delta S_S(\varphi=0)$
decreases with decreasing junction length and approaches the limit of
$\delta S_S(\varphi = 0)\rightarrow 0$ for $l\rightarrow 0$. The
temperature dependence of $\delta{S}_S(0)$ is largely affected by the
presence of a minigap in the spectrum,
$S(0)\sim{}e^{-E_g/T}$ with
$E_g\sim{}\min[\hbar{}D_N/L_N^2,|\Delta|]$ [see
Fig.~\ref{fig:schematics}(b)].  For $\varphi=\pi$ on the other hand,
the entropy contribution $\delta S_S$ of the superconductors increases
with decreasing length, in accordance with the increase of the
supercurrent with decreasing junction resistance. For very short
junctions, $\ell\lesssim{}a^{-1}$, $\delta S_S$ saturates as indicated
in Eq.~\eqref{eq:limits}.
The behavior of $\delta S_S (\varphi = \pi)$ at phase difference
$\varphi=\pi$ as a function of the product $al$ is shown in
Fig. \ref{fig:entropy_limits}(c). It is interesting to note that the
results are essentially converged to the short-junction limit $l\ll1$
already at $l=1$.

\begin{figure}
  \includegraphics{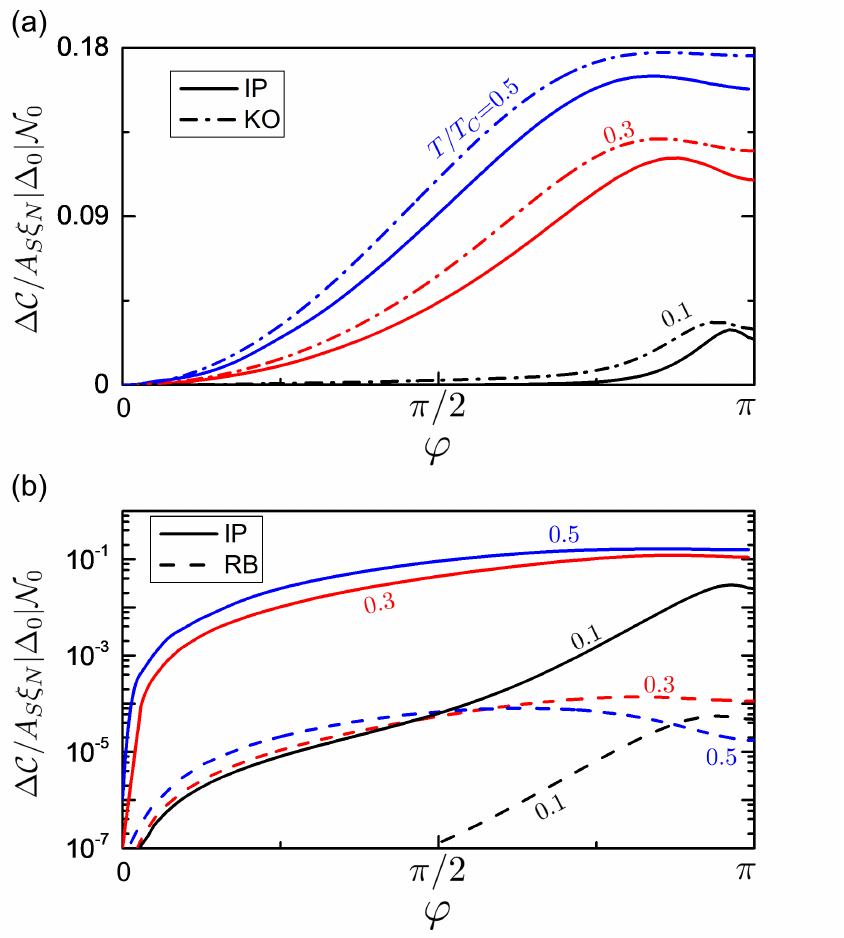}
  \caption{
    Modulation of the heat capacity $\Delta C(T,\varphi)=C(T,\varphi)-C (T,\varphi=0)$
    in an SNS junction, for $a=500$, $\ell=0.1$.
    (a)
    Numerical results including inverse proximity effect (IP, solid) and
    results from Eq.~\eqref{eq:KOentropy} (KO, dash-dotted),
    for different temperatures.
    (b)
    Same, shown on a logarithmic scale, together with
    the result from a rigid boundary condition approximation
    (RB, dashed).
  }
  \label{fig:heat}
\end{figure}

\section{Heat capacity}
\label{sec:heatcapacity}

The heat capacity
\begin{equation}
  C = T\frac{\dd{S}}{\dd{T}}
  \,,
\end{equation}
can be obtained from the entropy discussed in the previous
sections. Fig.~\ref{fig:heat} shows numerical results for the heat
capacity.  In these calculations, the order parameter $\Delta(x,T)$ is
computed to satisfy the self-consistency relations
$\delta{}F/\delta\Delta=\delta{}F/\delta\Delta^*=0$.  For the selected
short junction length, $al\gg1$, and the numerical results obtained by
taking the inverse proximity effect into account match relatively well
with Eq.~\eqref{eq:KOentropy}. Note that a
self-consistent $\Delta$ does not cause significant qualitative
deviations. On the other hand, calculations within the rigid boundary
condition approximation, shown in Fig.~\ref{fig:heat}(b),
underestimate the heat capacity by several orders of magnitude. As
pointed out above, we expect that this approach is accurate only for
long junctions $L_N\gtrsim{}5\xi_N$.

Finally, note that the total heat capacity at $\varphi=0$, being an
extensive quantity, will generally depend on device parameters of the
whole system.

\section{Summary and discussion}

\label{sec:concl}

The entropy in SNS junctions roughly consists of two contributions ---
a phase dependent part associated with the bound states contributing
also to the supercurrent, and a phase-independent part. Generally, the
two behave differently as a function of the junction length.
Moreover, the phase-dependent contribution in short junctions, if
expressed in terms of the local density of states, largely originates
from the proximity effect in the superconducting banks. Approximations
that neglect this can produce thermodynamically inconsistent results.
The results also reiterate, as clear from the connection to CPR, that the
junction heat capacity has a part not directly related to the junction
volume. A proper quantitative calculation of entropy and thermodynamic
quantities taking into account inverse proximity effect is thus of
importance both for fundamental and application purposes.

Finally, we can consider factors important for an experimental
measurement of the heat capacity of a single nanoscale SNS
junction. For example, the heat capacity of the junction can be inferred by measuring the temperature variation, after an heating pulse, as a function of the phase difference, which can be manipulated by means of external field. For such experimental realization, two points have to be considered with care. First, the device should be thermally well-isolated, in
order to avoid heat dispersion outside of device volume itself.
Second, the bulk superconductor mass should be made as small as
possible: the total heat capacity $C$ is an extensive property, so its
variation as a function of phase difference $\Delta C(\varphi)/C$
increases by increasing the ratio of critical current and device
volume.  However, this target will be also constrained by the
requirement of large superconducting leads in order to ensure the
phase-bias of the junction and thus an optimal trade-off has to be
considered in a proper device design.

In summary, we discussed entropy and heat capacity in SNS structures
numerically and analytically, and point out that inconsistencies
appear if inverse proximity contributions are not properly
included. The results obtained can be used in designing
superconducting devices concerning caloritronic, heat and photon
sensors, and are in general relevant also for other devices based on
thermodynamic working principles. 

\acknowledgments

We thank A. Braggio for discussions.  P.V. , F.V. and F.G. acknowledge
funding by the European Research Council under the European Union's
Seventh Framework Program (FP7/2007-2013)/ERC Grant agreement
No. 615187-COMANCHE and the MIUR under the FIRB2013 Grant
No. RBFR1379UX - Coca. 
M.C. acknoledges support from the CNR-CONICET cooperation programme ``Energy conversion in quantum, nanoscale, hybrid devices''.
The work of E.S. was funded by a Marie Curie
Individual Fellowship (MSCA-IFEF-ST No. 660532-SuperMag).
F. G. acknowledges funding by Tuscany Region under the FARFAS 2014 project SCIADRO.

\end{document}